Discussion Paper:Discussion Model for Propagation of Social Opinion via Quantum Galois Noise Channels:Entanglement, SuperSpreader(2023)

# From Spin States to Socially Integrated Ising Models: Proposed Applications of Graph States, Stabilizer States, Toric States to Opinion Dynamics

Yasuko Kawahata [†]

Faculty of Sociology, Department of Media Sociology, Rikkyo University, 3-34-1 Nishi-Ikebukuro,Toshima-ku, Tokyo, 171-8501, JAPAN.

ykawahata@rikkyo.ac.jp,kawahata.lab3@damp.tottori-u.ac.jp

**Abstract:** Recent research has developed the Ising model from physics, especially statistical mechanics, and it plays an important role in quantum computing, especially quantum annealing and quantum Monte Carlo methods. The model has also been used in opinion dynamics as a powerful tool for simulating social interactions and opinion formation processes. Individual opinions and preferences correspond to spin states, and social pressure and communication dynamics are modeled through interactions between spins. Quantum computing makes it possible to efficiently simulate these interactions and analyze more complex social networks.Recent research has incorporated concepts from quantum information theory such as Graph State, Stabilizer State, and Surface Code (or Toric Code) into models of opinion dynamics. The incorporation of these concepts allows for a more detailed analysis of the process of opinion formation and the dynamics of social networks. The concepts lie at the intersection of graph theory and quantum theory, and the use of Graph State in opinion dynamics can represent the interdependence of opinions and networks of influence among individuals. It helps to represent the local stability of opinions and the mechanisms for correcting misunderstandings within a social network. It allows us to understand how individual opinions are subject to social pressures and cultural influences and how they change over time.Incorporating these quantum theory concepts into opinion dynamics allows for a deeper understanding of social interactions and opinion formation processes. Moreover, these concepts can provide new insights not only in the social sciences, but also in fields as diverse as political science, economics, marketing, and urban planning.

## 1. Introduction

The study of approximate quantum algorithms for Ising distribution functions began in physics, particularly in the field of statistical mechanics. The Ising model is a simple mathematical model developed to understand the physical properties of magnetic materials and subsequently applied to a variety of scientific and engineering problems. The model is designed to simulate the behavior of spin systems, where the interaction of spins can generate complex patterns and phase transitions. In recent years, this model has played an important role in the field of quantum computing, particularly in quantum annealing and quantum Monte Carlo methods.

In opinion dynamics, the Ising model is also a powerful tool for simulating social interactions and opinion formation processes. In this field, an individual's opinions and pref-erences correspond to spin states, and social pressure and communication dynamics are modeled through interactions between spins. Quantum computing makes it possible to efficiently simulate these interactions and analyze more complex social networks.

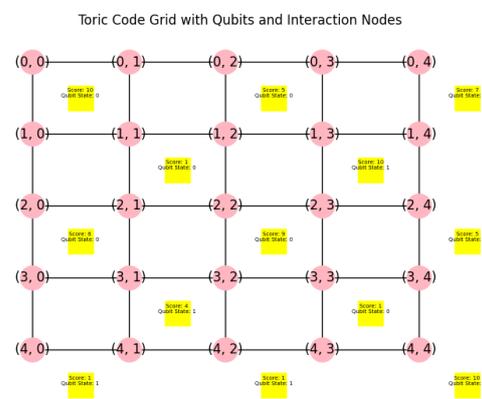

Fig. 1: Tile Graph of Interaction Coefficients:Quantum Toric Code Opinion Dynamics

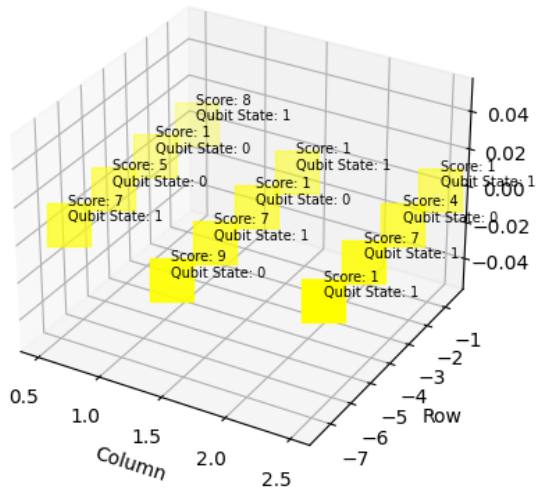

Fig. 2: 3D Map of Interaction Coefficients:Quantum Toric Code Opinion Dynamics:1

Recent research has incorporated concepts from quantum information theory such as Graph State, Stabilizer State, and Surface Code (or Toric Code) into models of opinion dynamics. The incorporation of these concepts allows for a more detailed analysis of the process of opinion formation and the dynamics of social networks.

Graph State is a concept that lies at the intersection of graph theory and quantum theory, where nodes correspond to qubits (cubits) and edges correspond to entanglements between cubits. Graph State can be used in opinion dynamics to represent the interdependence of opinions and networks of influence among individuals.

Stabilizer States play an important role in quantum error correction, but in opinion dynamics they are used to represent the stability and consistency of opinions. It allows us to analyze how individual and collective opinions evolve over time and what factors cause these changes.

Surface Code (Toric Code) is a two-dimensional lattice model used for error correction of quantum information. In opinion dynamics, it helps to represent the local stability of opinions and the mechanisms for correcting misunderstandings within a social network. This allows us to understand how individual opinions are affected by social pressures and cultural influences, and how they change over time.

and how they change over time.

Incorporating these quantum theory concepts into opinion dynamics allows for a deeper understanding of social interactions and opinion formation processes. Moreover, these concepts can provide new insights not only in the social sciences, but also in fields as diverse as political science, economics, marketing, and urban planning. With the advancement of quantum computing technology, these approaches are expected to develop further in the future and open up new areas of social science.

## 2. Ising Models, Quantum Complexity Research

### 2.1 Case Studies in Quantum Computing

These papers present important research related to quantum computing and information theory. Following is a summary of what each paper features and what research it has done: Preskill (1998) provided comprehensive lecture notes on quantum computation and information in Quantum Computation and Information. This paper details the basic principles and concepts of quantum computation. In "Algorithms for quantum computation: discrete logarithms and factoring," Shor (1994) proposed algorithms for discrete logarithms and factorization in quantum computation. These algorithms demonstrate the ability of quantum computers to solve these problems faster than classical computers.

Grover (1996) introduced a fast quantum mechanical algorithm for database search in "A fast quantum mechanical algorithm for database search. This algorithm has been shown to solve search problems faster than classical algorithms.Deutsch (1985), in "Quantum theory, the Church-Turing principle and the universal quantum computer," presented theoretical considerations. This paper explores in depth the relationship between quantum and classical computation. Aaronson (2007) provided a discussion of the limits of quantum computers in "The Limits of Quantum Computers. The paper discusses extensively the possibilities of quantum computation and its limitations, focusing on open problems. These papers represent important contributions to the field of quantum computing and information theory. They focus on a wide range of topics, from fundamental principles of quantum computation to fast algorithms and theoretical considerations. They provide insights into the possibilities and limitations of quantum computing and into future information processing technologies.

### 2.2 Ising Models and Computational Complexity Research

These papers present important research on Ising models and computational complexity. The following is a summary of the features of each paper and the type of research it has conducted:

Barahona (1982) conducted a study of the computational complexity of Ising spin glass models in "On the computational complexity of Ising spin glass models". This paper explored theoretically the difficulty of computational problems in Ising models.

Lucas (2014), in "Ising formulations of many NP problems," proposed a method to convert many NP problems into Ising models. This clarified the connection between computational complexity theory and Ising models.

Park and Newman (2013) provided a solution to the Ising problem on a chimera graph in "Solution of the Ising problem on a chimera graph". This is related to the development of efficient solution methods for specific hardware structures.

Lucas (2018) described the development of Ising machines over the past 70 years in "Ising machines: the first 70 years". The paper focuses on different approaches, including classical computation and quantum annealing.

Megow and Verschae (2016), in "Solving the Ising problem with a D-Wave quantum annealer," proposed a way to solve the Ising problem using D-Wave quantum annealing machines. This is an important study on the practicality of quantum computing.

These papers present important research at the intersection of Ising models and computational complexity theory. Ising models can be applied to a wide variety of computational problems and their computational complexity is extensive. These studies are relevant to both classical computation and quantum computing and point to important advances in the field of computational science.

## 2.3 Quantum Annealing and Physics Research

These papers present important research focused on the connection between quantum annealing and physics. Below is a summary of what each paper features and what research it has done: In "Quantum Annealing for the Number-Partitioning Problem," Pudenz and Lidar (2013) used quantum annealing to address the number partitioning problem. This work was one of the early attempts to apply quantum computing to optimization problems. Boixo et al. (2014) provided evidence for quantum annealing with more than 100 qubits in "Evidence for quantum annealing with more than one hundred qubits." This study provided insight into the scalability of quantum annealing. Katzgraber and Young (2015), in "Monte Carlo methods in the physical sciences: celebrating the 50th anniversary of the Metropolis algorithm" They highlighted the importance of Monte Carlo methods in the physical sciences. Monte Carlo methods are widely used in the physical and computational sciences. Albash and Lidar (2018) provided a comprehensive review on slow quantum annealing (computation in insulating quantum systems) in "Adiabatic quantum computation". The paper provides a detailed description of the basic principles of quantum annealing. Kadowaki and Nishimori (1998) studied quantum annealing in the trans-verse Ising model in "Quantum annealing in the transverse Ising model. This paper was an important step in the understanding of quantum annealing in the framework of statistical and quantum mechanics. These papers provide important insights into the interface between quantum annealing and physics. Quantum annealing has been applied to a variety of problems, including optimization and combinatorial optimization, and research is ongoing on both its theoretical and practical aspects. Understanding the relationship between physics methods and quantum annealing has also contributed to the development of quantum computing.

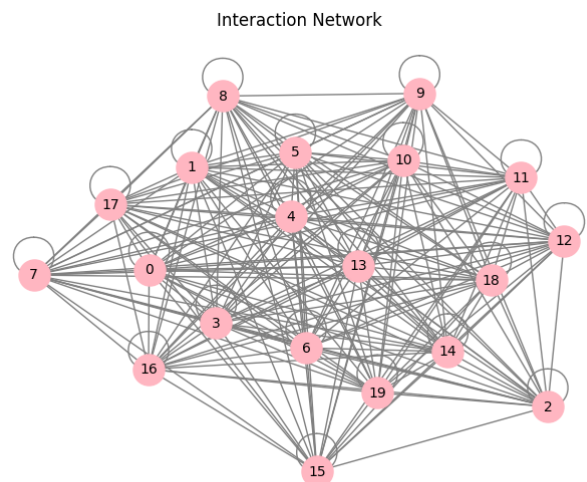

Fig. 3: Quantum Opinion Dynamics Network

# 3. Method of Approximate Quantum Algorithms for the Ising Distribution Function

The following is an introduction to the Opinion Dynamics research paper. We will cover the historical background of research related to approximate quantum algorithms for Ising distribution functions, which will then be used to introduce the importance of the study of Opinion Dynamics.

Opinion formation and its diffusion in social networks is regarded as an important research topic in modern society. Understanding how individuals' opinions and beliefs spread and influence each other within networks provides important insights in a variety of fields, including politics, marketing, information transfer, and even epidemiology. This phenomenon, called opinion dynamics, is driven by a complex network of interactions and information transfer between individuals, and many mathematical models and computational methods have been proposed to understand it.

In this paper, we focus on the application of approximate quantum algorithms for the Ising distribution function to the study of opinion dynamics and its new insights and results.

We have focused on exploring possibilities beyond classical models and methods, and on harnessing the power of quantum computation to better understand the mechanisms of opinion formation and diffusion within social networks. The results of this research are expected to have far-reaching implications for a wide range of applications related to opinion formation, information transfer, and social dynamics.

Opinion formation and its diffusion in social networks is regarded as an important research topic in contemporary society. Understanding how individuals' opinions and beliefs spread and influence each other within networks provides important insights in a variety of fields, including politics, marketing, information transfer, and even epidemiology. This phenomenon, called opinion dynamics, is driven by a complex network of interactions and information transfer between individuals, and many mathematical models and computational methods have been proposed to understand it.

In addition, the incorporation of the concept of Surface Codes (or Toric Codes) in the modeling of opinion dynamics is opening new vistas. These codes are methods that utilize a two-dimensional lattice-like structure for quantum error correction, and their distinctive properties may be applied to the study of opinion dynamics to model the local stability of opinions and error (misinformation or misunderstanding) correction mechanisms within social networks.

Research on approximate quantum algorithms for Ising distribution functions originally started in statistical physics, but is now considered very important in the field of quantum computing. This approach is used to simulate particle interactions and its applications range from physics to chemistry, biology, economics, and even the social sciences. Specifically, it is used to analyze opinion dynamics, or the process by which people form opinions. In this field, Ising models are used to simulate individual opinions and social interactions, and quantum computing is used to efficiently analyze these dynamics.

The Ising model was originally developed to understand the magnetization of magnetic materials, but has since been applied to the study of more complex interacting systems. In quantum computing, particularly quantum annealing and quantum Monte Carlo methods, the model has contributed to increased computational efficiency. The use of qubits (cubits) allows efficient exploration of large state spaces that cannot be handled by conventional computers.

In the field of opinion dynamics, Ising models play an important role in modeling the influence of opinions among individuals and social pressures. Individual opinions are treated as spins, and the interaction of these spins with each other simulates the change or diffusion of opinions. By incorporating quantum computing techniques, these processes can be simulated with greater speed and accuracy. More recently, the study of opinion dynamics has incorporated concepts from quantum information theory such as Graph State, Stabilizer State, and Surface Code (or Toric Code). By incorporating these concepts into models, the process of opinion formation and the dynamics of social networks can be analyzed in greater detail. Graph State is a concept that links graph theory and quantum theory, where nodes represent cubits and edges represent entanglements between cubits. In opinion dynamics, it is used to represent the interdependence of opinions and influence networks among individuals. Stabilizer State is important in quantum error correction, but in opinion dynamics it is used to represent the stability or consistency of opinions. This allows us to analyze how individual and collective opinions change over time and what causes these changes and what causes those changes. Surface Code is a two-dimensional lattice model used for error correction of quantum information. In opinion dynamics, it helps to describe the local stability of opinions and the mechanisms for correcting misunderstandings in social networks. It allows us to understand how individual opinions are affected by social pressures and cultural influences and how they change over time.

Incorporating these quantum theory concepts into opinion dynamics allows for a deeper understanding of social interactions and opinion formation processes. In addition, these concepts are expected to bring new insights not only to the social sciences, but also to fields as diverse as political science, economics, marketing, and urban planning. As quantum computing technology advances, these approaches are expected to be further refined and to open up new areas of social science.

## 3.1 Idea of Quantum Algorithms for the Ising Distribution Function

Research on approximate quantum algorithms for Ising distribution functions began in physics, particularly in statistical mechanics, and has become an important application area in quantum computing. The Ising model is a mathematical model developed to understand physical phenomena such as the magnetism of matter, which simulates the interaction of spins. The model has applications in a wide range of fields besides physics, including chemistry, biology, economics, and social sciences.

The Ising distribution function describes the statistical properties of these spin systems, but it is usually very difficult to compute this function accurately. For this reason, the development of approximation algorithms is of great importance, and quantum computer-based approaches have attracted much attention. Specifically, quantum annealing and quantum Monte Carlo methods are being used to compute the properties of Ising models faster or more accurately than traditional computing methods.

The approach with quantum annealing, the process of

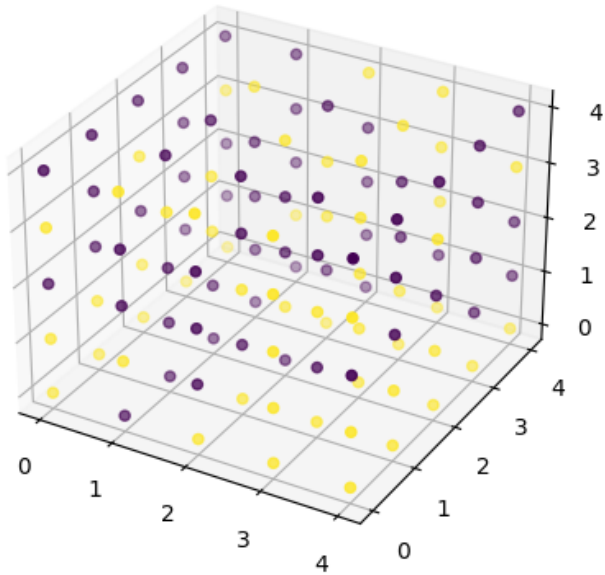

Fig. 4: 3D:Tile Graph of Interaction Coefficients Quantum Opinion Dynamics

## 3.2 Approximate quantum algorithm

Consider applications to the Stabilizer State, Graph State, and Surface Code (or Toric Code) State related to the approximate quantum algorithm for the Ising distribution function.

### 1.Stabilizer State

Definition: Stabilizer States refer to specific quantum states associated with quantum error-correcting codes. These states are defined by a set of stabilizers and are used for error correction and protection of quantum information. Visualization Method: Stabilizer states are often represented as a series of stabilizer operators, which are shown in mathematical notation. While intuitive graphical rendering is difficult, it is possible to represent them as quantum circuits and visualize them as vectors of quantum states.

### 2. Graph State

Definition: A Graph State is a quantum state constructed based on a particular graph structure. Here, nodes represent cubits and edges represent entanglement between cubits. Visualization Method: Graph States can be directly visualized using the corresponding graph structure. Nodes and edges can be used to show the interaction and entanglement relationships between cubits.

### 3. Surface Code (Toric Code) State

Definition: Surface Code is a method of quantum error correction based on a two-dimensional lattice-like structure. Visualization method: They are visualized as a 2-dimensional or 3-dimensional lattice, with the cue bits placed at the vertices or edges of the lattice. The error correction process can be visualized on this lattice.

These visualization methods will facilitate a deeper understanding of quantum states and help to clearly illustrate complex concepts and processes in quantum computing.

Here is the translated text with LaTeX code for the formulas and parameters: It is theoretically possible to model opinion dynamics (opinion formation dynamics) by applying an approximate quantum algorithm for the Ising partition function. Opinion dynamics model the process of opinion formation and dissemination within social networks, and the Ising model provides a suitable framework for representing such systems.

## 4. Opinion Dynamics:Ising Model

In the Ising model, each particle (spin) can be in one of two states: "up" or "down". When applied to opinion dynamics, the state of each particle can represent an individual's opinion (for example, "for" or "against"). Interactions between particles model the influence of opinions between individuals.

mapping the Ising model to qubits (cubits) and gradually transforming them from an initial state into a Hamiltonian that represents the model. This approach tracks changes in the quantum state while maintaining a state close to the ground state, eventually sampling the state of the Ising model. For example, D-Wave Systems' quantum annealing machine uses this approach and is being investigated for applications in a variety of fields, including optimization problems, machine learning, and materials science.

Another approach is the quantum Monte Carlo approach, which applies a specific quantum gate to a cubit system to stochastically sample the state space of the system. This method is used to estimate the probabilities for different terms of the distribution function of the Ising model. Applications of quantum Monte Carlo methods include the study of changes of state of matter and chemical reactions, as well as financial modeling.

These quantum algorithms have the potential to provide better results than classical approaches, especially in large Ising models and complex interacting systems. However, the current level of technology and error rates of quantum computers must be taken into account. Quantum algorithms continue to evolve and new techniques and ideas will be introduced regularly.

## Formulas and Parameters

The Hamiltonian of the Ising model is expressed as follows:

$$H = \sum_{\langle i,j \rangle} J_{ij} s_i s_j \sum_i h_i s_i$$

where, $s_i$ is the spin (or opinion) of the particle (or individual) $i$. $J_{ij}$ represents the strength of interaction between particles $i$ and $j$ (reflecting the magnitude of influence between individuals). $h_i$ is the external field (which can represent external factors or media influence).

## Application to Opinion Dynamics

1. Network Construction: Construct a network with individuals as nodes and their social connections as edges.

2. Setting Interactions: Set the influence between individuals ($J_{ij}$), based on the strength of relationships and degree of opinion influence.

3. Setting External Fields: Use $h_i$ to model the influence of external media and other factors.

4. Simulation: Evolve the system according to the dynamics of the Ising model, and simulate changes and distributions of opinions.

Quantum algorithms can be used to simulate the dynamics of this Ising model. Specifically, quantum annealing and quantum Monte Carlo methods are effective for large systems or systems with complex interactions. Modeling opinion dynamics using the Ising model helps deepen our understanding of opinion formation and dissemination in social networks. Additionally, the use of quantum algorithms offers the potential for more efficient simulation of such complex dynamics. However, detailed data and sophisticated modeling are required to accurately model realworld social networks.

# 5. Discussion

## Stabilizer States

Incorporating the conditions of Stabilizer States into the modeling of opinion dynamics signifies applying concepts from quantum information theory within the context of social sciences. Stabilizer States provide a framework for describing quantum states in certain quantum error correction codes. To apply this to opinion dynamics, it's necessary to translate these abstract concepts into concrete social science models.

## Basics of Stabilizer Formalism

In Stabilizer Formalism, quantum states are characterized by stabilizer operators. These operators "stabilize" a specific quantum state, meaning they are operators that do not change that state. In a qubit system, stabilizers are described as elements of the group generated by Pauli matrices.

## Application to Opinion Dynamics

When considering Stabilizer States in the context of opinion dynamics, each individual's opinion state is modeled as a qubit, and stabilizer operators are interpreted as rules that "stabilize" these opinion states. This can be used to represent how an individual's opinion is influenced by specific social pressures or norms.

1. Representation of Opinion States: Represent the opinion of individual $i$ as a qubit $q_i$. For instance, $|0\rangle$ might represent one opinion (e.g., in favor), and $|1\rangle$ another opinion (e.g., against).

2. Definition of Stabilizer Operators: Define a stabilizer $S_i$ as an operator that acts on the opinion state $q_i$ and does not change it. For example, consider $S_i$ such that $S_i q_i = q_i$.

3. Physical Interpretation of Stabilizers: The stabilizer $S_i$ models how an individual's opinion is "stabilized" by social networks, media, and cultural norms. For instance, $S_i$ can be set as a parameter representing the influence from a particular social group or information source.

4. Modeling Interactions: Interactions between individuals can also be considered. For example, the interaction between two individuals $i$ and $j$ can be modeled using a stabilizer $S_{ij}$. This interaction can represent the influence of opinions between individuals. Parameters of Stabilizers: Define the specific form of the stabilizer $S_i$ (e.g., a combination of Pauli matrices) and how it affects the opinion state. Strength of Interaction: The strength of the interaction between individuals is defined through the stabilizer $S_{ij}$, reflecting the strength of relationships and the degree of opinion sharing.

## Caveats

This approach is an example of applying quantum information theory concepts to social sciences. The direct correlation to actual social dynamics is not always clear. Care is needed in interpreting quantum theory concepts in the context of social sciences. Concepts like qubits and stabilizers may function as metaphors in social science models. This framework has the potential to offer new perspectives in the study of opinion dynamics, but careful consideration is required in its interpretation and application.

In quantum computation, rather than taking on the state of 0 or 1, a bit can take on both 0 and 1 states simultaneously in the form of a superposition. The Bloch sphere is used to visualize this superposition state. Using randomly generated theta (theta) and phi (phi) angles, each qubit is initialized to a specific superposition state. Although this process is different from social phenomena and opinion dynamics, the superposition of qubits can be thought of as analogous to representing the spectrum of social opinion as a continuum rather than a simple binary choice (0 or 1). That is, opinions on an issue can be diverse, not simply for or against, but also taking intermediate positions. The fact that the state of each qubit is in a different position on the Bloch sphere indicates the diversity of opinion states held by different individuals. These qubits can form quantum entanglement through interaction, which can also be thought of as representing the complex interactions between individuals in social opinion formation. Also, when there is interaction between qubits, it is represented by the formation of quantum entanglement, which can be seen as a quantum analogue of how opinions between individuals influence each other. In quantum computation, "stabilizer operators" are used as operators that preserve a particular quantum state, but such concepts do not directly apply in this context. However, the fact that the state of a qubit is initialized by random parameters can be viewed as a process by which an individual forms its opinion based on random elements.

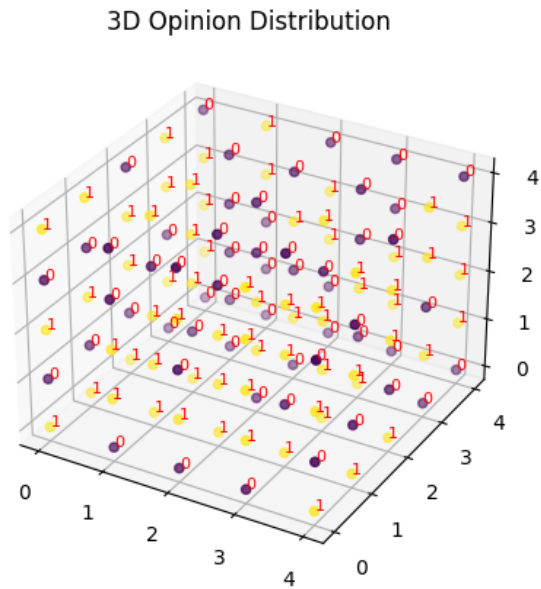

Fig. 5: Quantum Stabilizer State Opinion Dynamics

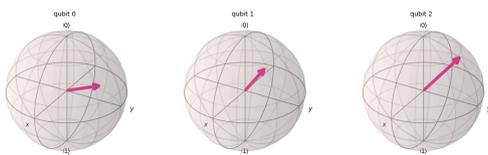

Fig. 6: Quantum Stabilizer State Opinion Dynamics

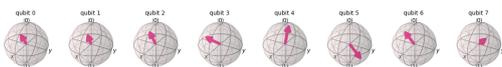

Fig. 7: Quantum Stabilizer State Opinion Dynamics

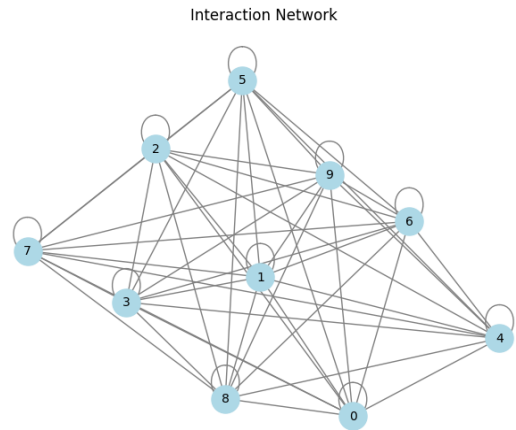

Fig. 8: Quantum Stabilizer State Opinion Dynamics

In an actual quantum computation, these qubits play the role of bits in a computational task, but by incorporating them as a metaphor in a social science context, we may be able to examine social phenomena and the dynamics of opinion formation from a new perspective.

Fig. 6 shows the interaction network and how individuals are interconnected. This models a social network or relationship between people. The structure of this network suggests how opinions tend to spread or how certain individuals may play a central role in opinion formation.

The mesh image of opinions and interactions shows the complex patterns of how opinions interact among individuals. From this image, one can read how opinions are spatially concentrated and how they are dispersed. Darker colored areas probably indicate areas where opinions are closely interacting, suggesting that clusters of opinions are forming. The density of the interaction network also indicates the strength of the interaction between individuals within the network. A strongly interacting network indicates an environment in which opinions can propagate quickly and can indicate a tendency toward consensus building or polarization of opinions. The stabilizer operator image is an indicator of how stable (i.e., hard to change) each individual's opinion is. Darker colored regions may indicate individuals whose opinions are more stable. This distribution helps us understand how opinions may change over time.

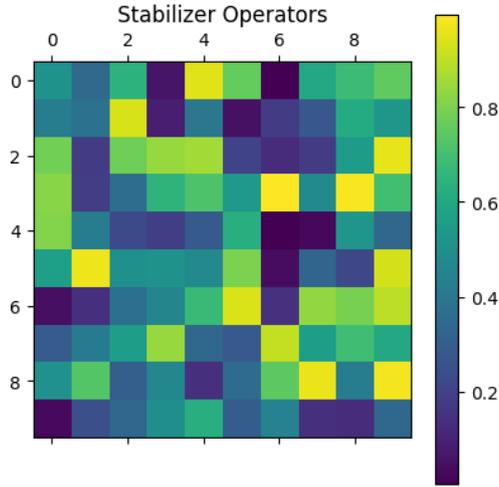

Fig. 9: Quantum Stabilizer State Opinion Dynamics:Tile Graph of Interaction Coefficients

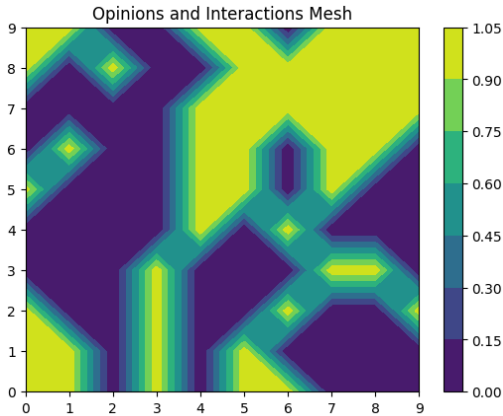

Fig. 10: Quantum Stabilizer State Opinion Dynamics:Mesh Graph of Opinions and Interactions

This model is useful for understanding the process of social opinion formation and may be particularly applicable to social network analysis and the study of collective behavior. Of course, comparisons with actual data are necessary to assess how well these models reflect reality in actual social phenomena.

# Graph States

Incorporating the concept of Graph States into the modeling of opinion dynamics represents a highly innovative approach of applying the framework of quantum information theory to the context of social sciences. Below is a general proposal for expressing this idea mathematically.

## 5.1 1. Definition of Qubits:

Represent the opinion of individual $i$ with a qubit $q_i$. $q_i$ can take the base states $|0\rangle$ or $|1\rangle$. The state of the qubit can represent an individual's opinion, such as 'for' or 'against'.

## 5.2 2. Generation of Graph States:

Construct a graph $G$ and correspond each node to a qubit, with edges representing interactions between qubits. An edge $(i, j)$ can represent the entanglement between qubits $q_i$ and $q_j$.

## 5.3 3. Definition of Hamiltonian:

Define the system's Hamiltonian $H$ as follows:

$$H = \sum_{(i,j) \in G} J_{ij} \sigma_z^{(i)} \sigma_z^{(j)} + \sum_i h_i \sigma_x^{(i)}$$

Here, $J_{ij}$ represents the coefficient of interaction between qubits, $h_i$ represents the coefficient of external field influence, and $\sigma_z$ and $\sigma_x$ are Pauli matrices.

## 5.4 4. Dynamics of the System:

The time evolution of the system can be simulated using the Schrödinger equation or quantum circuits. The creation and manipulation of entanglement can be achieved using CNOT gates and other quantum gates.

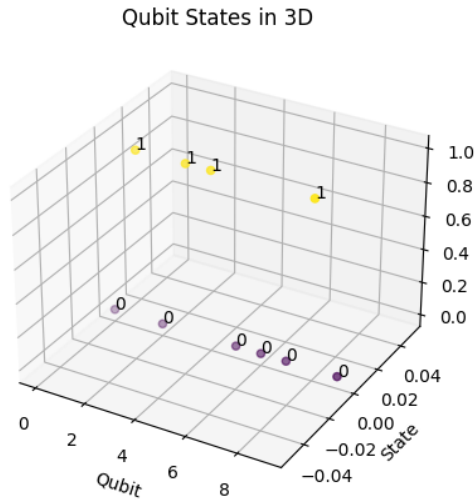

Fig. 11: Quantum Graph State Opinion Dynamics

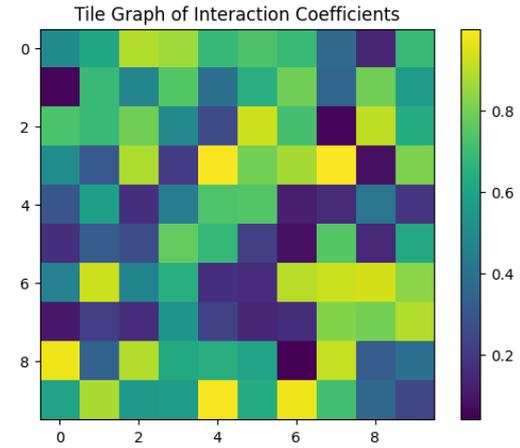

Fig. 13: Tile Graph of Interaction Coefficients:Quantum Graph State Opinion Dynamics

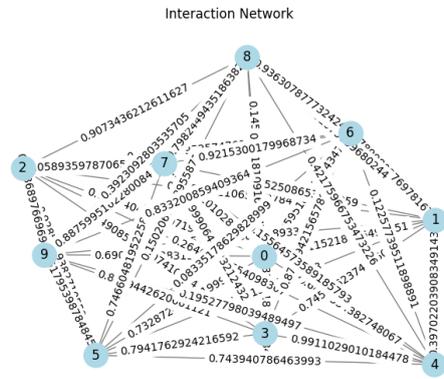

Fig. 12: 3D:Quantum Graph State Opinion Dynamics

**Setting Parameters**

Interaction Coefficient $J_{ij}$: Represents the strength of interaction between qubits and models the influence of opinions between individuals. External Field Coefficient $h_i$: Represents external factors influencing an individual's opinion (e.g., media or cultural influences).

## 5.5 Caveats

This model includes metaphorical interpretations and does not indicate a direct physical correspondence. Additional data and experimental validation are required to model the dynamics of opinions in actual social networks. This approach provides a theoretical framework for exploring new interactions between quantum information theory and social sciences. In practical research and applications, it is important to assess the effectiveness and limitations of this model through concrete data and experimental validation.

The states of the qubits (cubits) and the interaction coefficients (entanglement) between them are visualized. By viewing the states of the qubits as social opinions, we can think of these qubits as representing individuals exhibiting simple binary opinions. In this simulation, individuals have an opinion of 0 or 1. As a consideration, note that in real social phenomena, opinions are more continuous and complex. The interaction coefficient between each cue bit can be considered as the influence between opinions. It represents the strength of the social interaction, which is used to understand how opinions interact and evolve over time. The first 3D graph shows the distribution of opinions by cubit, and in this representation it appears as if the opinions of each cubit are independent. In reality, however, their states are dependent on each other due to entanglement among the cubits. In the context of social science, this is equivalent to modeling how individuals are affected by each other within a group. The graph of the interaction network shows the interaction coefficient $J_{ij}$ between the cubits. The larger this coefficient is, the stronger the entanglement between the two cubits. This corresponds to the strength of the ties between individuals in a social network and is a factor that affects the speed and pattern of opinion propagation.

Such a model provides a theoretical framework for studying the dynamics of opinion formation within social groups. Of course, it is important to test the validity of the model using actual social network data.

Considerations based on a model that visualizes the states of qubits and the interaction coefficients between them. The tile graph represents the strength of the interaction between individual agents (qubits) in a social network. Light tiles indicate strong interactions and dark tiles indicate weak interactions. This pattern reflects the degree of social connectedness and influence, for example, the relationship between opinion leaders and their followers, and possibly the existence of a closer-knit community.

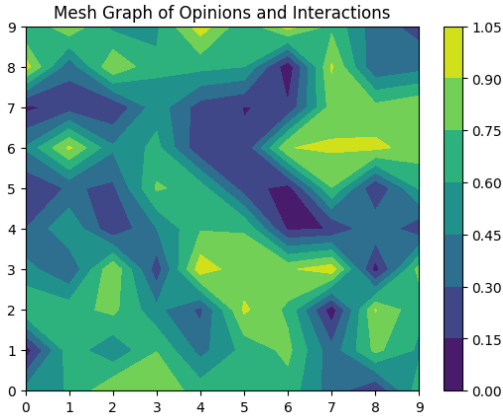

Fig. 14: Mesh Graph of Opinions and Interactions:Quantum Graph State Opinion Dynamics

Considered as opinion dynamics, the mesh graph represents the differences in opinion and strength of interaction between individuals in a continuous space. This visualization can represent the concentration or dispersion of opinions, or patterns of opinion flow or change, helping to understand social opinion dynamics.

As a discussion of the distribution of opinions by cubit, the binary distribution of opinions represented by a cubit is shown. This represents the balance of opinion in a group at a given point in time, and this information helps us understand the diversity of opinion and whether one opinion is dominant or not.

Discussion of the behavior of the interaction coefficient $J_{ij}$ by entanglement**:. - The interaction coefficient $J_{ij}$ between tile and mesh graphs represents the degree of entanglement between cubits. High entanglement, in a social context, implies strong social influence and close relationships, which play an important role in opinion propagation and change.

# 6. Conclusion
# Toric State(Code)

Incorporating the concept of the Toric Code into the modeling of opinion dynamics is a complex endeavor that further develops this modeling. The Toric Code is a method that utilizes a 2D lattice structure for quantum error correction. Applying this to opinion dynamics might allow us to model the local stability of opinions within social networks and the mechanisms for correcting errors (such as misinformation or misunderstandings).

## Application of Toric Code to Opinion Dynamics

The key features of the Toric Code are qubits arranged on a 2D lattice and specific patterns of error correction operators to stabilize these qubits. This can be applied to opinion dynamics as follows.

### 1. Definition of Qubits

Place qubits $q_{ij}$ representing individual opinions at each point on a 2D lattice. Each qubit, for instance, could represent one opinion with $|0\rangle$ and another opinion with $|1\rangle$.

### 2. Introduction of Error Correction Operators

At the core of the Toric Code are error correction operators to stabilize the states of qubits. These can be interpreted as social mechanisms to 'correct' misinformation or misunderstandings.

### 3. Definition of Hamiltonian

The Hamiltonian $H$ of opinion dynamics can include terms representing interactions between individuals and influences from external sources.

Interactions Between Qubits: Interactions between adjacent points on the lattice represent the influence and sharing of opinions. This is defined by interaction coefficients $J_{ij}$. Error Correction Mechanisms: Model mechanisms for correcting misinformation or misunderstandings through error correction operators. These operators can represent social pressures, norms, education, and other mechanisms for correcting misinformation.

## Caveats

The concept of the Toric Code, originally from the context of quantum error correction, becomes highly abstract when applied to social sciences, requiring careful interpretation. While this model has the potential to enrich the theoretical framework of opinion dynamics, the direct correlation with actual social dynamics might not be clear. When applying quantum theory concepts to social sciences, it's important to clearly understand the difference between their roles as metaphors and their physical reality. The application of the Toric Code to opinion dynamics could provide new insights into the flow of information and correction mechanisms for misunderstandings within social networks.

## (1)Consideration as a Social Phenomenon

The use of qubits and their interactions to model social phenomena is quite innovative. Qubits allow for the representation of states that are not just binary (as in classical models), but can also exhibit superposition and entanglement, which could metaphorically represent the complex and sometimes non-deterministic nature of human opinions and social interactions. For instance, the state of a qubit could represent an individual's opinion, which is not just a simple yes/no but can be a complex combination of factors.

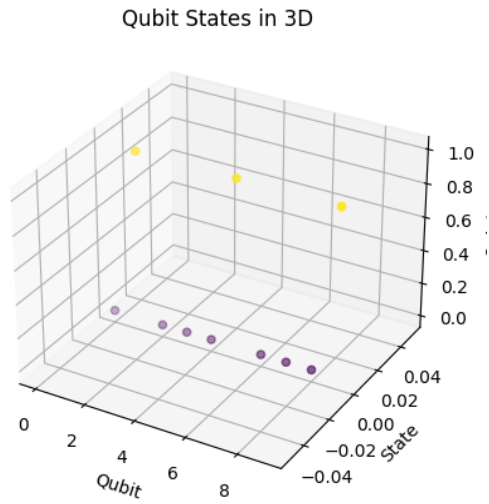

Fig. 15: 3D Map of Interaction Coefficients:Quantum Toric Code Opinion Dynamics:1

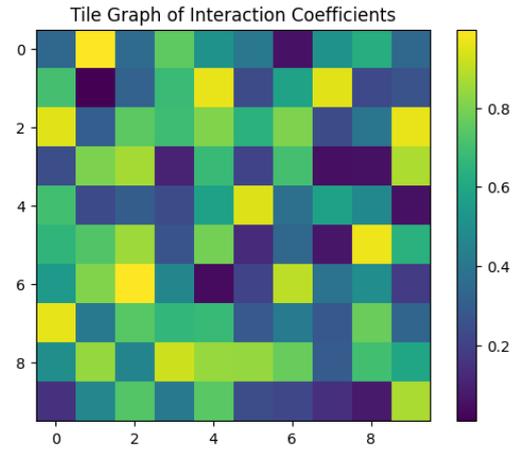

Fig. 17: Tile Graph of Interaction Coefficients:Quantum Toric Code Opinion Dynamics

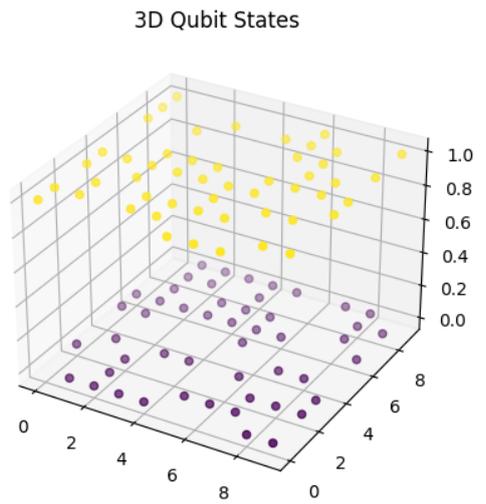

Fig. 16: 3D Map of Interaction Coefficients:Quantum Toric Code Opinion Dynamics:1

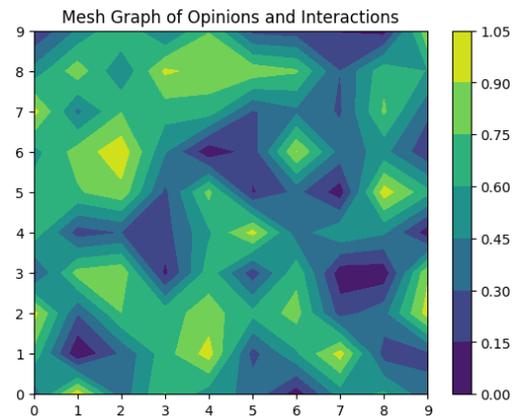

Fig. 18: Mesh Graph of Opinions and Interactions:Quantum Toric Code Opinion Dynamics

## (2) Consideration of Opinion Dynamics

Opinion dynamics often deal with how individual opinions change over time due to factors like personal conviction and social influence. In this quantum-inspired model, the interaction coefficients $J_{ij}$ could represent the strength and nature of influence between individuals $i$ and $j$, similar to coupling constants in physical systems. The dynamics could then be analyzed by how these coefficients affect the state of the qubits over time.

## (3) $q_{ij}$ Distribution of Opinions for Different Qubits

The code generates a random distribution of qubit states, which can represent a diverse set of opinions within a population. In a quantum system, the distribution of these states can be influenced by interactions with other qubits, suggesting how a consensus or disagreement may form in a social group.

## (4) $J_{ij}$ Interaction Coefficients by Entanglement

Entanglement is a uniquely quantum phenomenon where the states of two qubits become linked, such that the state of one cannot be described independently of the state of the other. In social models, this could represent deep connections between individuals where the opinion of one strongly influences the other. The interaction coefficients in an entangled state could be very high, reflecting this strong connection.

## (5) $h_{ij}$ Behavior of External Field Influence $h_{ij}$)

In physics, an external field can change the state of a system. In social terms, this could represent external media influence or cultural norms that affect individual opinions. The model could include this by having an external field term $h_{ij}$ that influences the state of the qubits. This would allow the model to consider both internal interactions and external influences.

The tile graph shows interaction coefficients, which could indicate how strongly each pair of qubits (individuals) influences one another. The mesh graph appears to combine these coefficients with the states, giving a contour map of opinions and interactions.

For a more detailed analysis, we would need to know how the interaction coefficients and the qubit states are updated in time (the dynamics), and how the external field is applied to the system.

This structure can also be used metaphorically to represent a network of individuals (qubits) with their opinions (qubit states) and the interactions (edges) between them, potentially influenced by an external field (environmental factors).

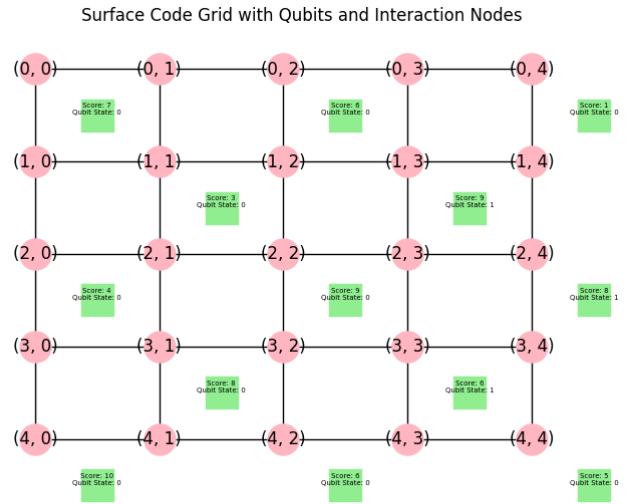

Fig. 19: Tile Graph of Interaction Coefficients: Quantum Toric Code Opinion Dynamics

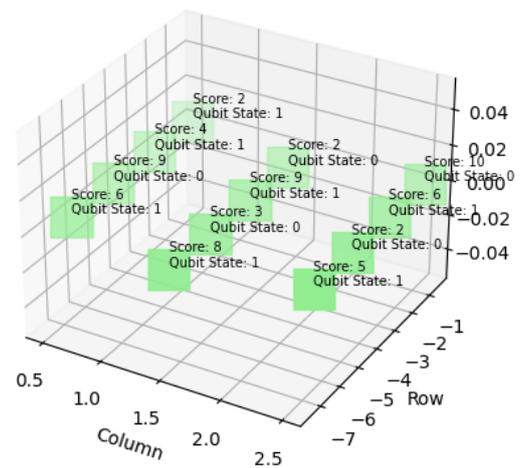

Fig. 20: 3D Map of Interaction Coefficients: Quantum Toric Code Opinion Dynamics:1

### (1)Consideration as a Social Phenomenon

The surface code grid can represent a social network where nodes are individuals and edges are the relationships or pathways through which opinions can influence one another. The "Score" could represent the strength or intensity of an individual's opinion, and the "Qubit State" might indicate the stance (positive, neutral, negative). This grid can model the complexity of social interactions and how opinions spread and evolve over time.

### (2)Consideration of Opinion Dynamics

The qubit states with associated scores could represent how strongly individuals hold their opinions. The dynamics of these opinions can be studied by examining how changes in one node (individual) affect others. In a quantum analogy, this could relate to the concept of superposition, where an opinion might be a combination of states until an interaction (measurement) causes it to collapse to a definite state.

### (3) $q_{ij}$ Distribution of Opinions for Different Qubits

The distribution of opinions across the grid reflects the diversity of a social group. This distribution could also indicate clusters of consensus or disagreement. For example, areas with similar scores and qubit states might represent groups with shared opinions, while areas with diverse scores could indicate a contentious topic.

### (4) $J_{ij}$ Interaction Coefficients by Entanglement

In this analogy, entanglement could represent a scenario where the opinion of one individual cannot be fully described without considering another's opinion. High interaction coefficients might then represent relationships where opinions are strongly correlated or entangled. In social terms, this could be akin to close-knit communities or peer influence.

### (5) $h_{ij}$ Behavior of External Field Influence $h_{ij}$

The external field's influence can be likened to societal pressures, media influence, or cultural norms that can sway individual opinions. The behavior of $h_{ij}$ might illustrate how individuals or groups of individuals react to external stimuli, with some changing their opinions (state change) while others remain unaffected.

To advance the analysis, it would be helpful to understand the rules governing the state changes for the qubits, the exact nature of the scores, how the interaction coefficients are determined, and how the external field is quantified and applied within this model.

## Aknowlegement

The author is grateful for discussion with Prof. Serge Galam and Prof.Akira Ishii. This research is supported by Grant-in-Aid for Scientific Research Project FY 2019-2021, Research Project/Area No. 19K04881, "Construction of a new theory of opinion dynamics that can describe the real picture of society by introducing trust and distrust".

## References

zh

[1] Preskill, J. (1998). Quantum Computation and Information. *Caltech lecture notes*.

[2] Shor, P. W. (1994). Algorithms for quantum computation: discrete logarithms and factoring. In *Proceedings of the 35th Annual Symposium on Foundations of Computer Science*.

[3] Grover, L. K. (1996). A fast quantum mechanical algorithm for database search. *Proceedings of the twenty-eighth annual ACM symposium on Theory of computing*.

[4] Deutsch, D. (1985). Quantum theory, the Church-Turing principle and the universal quantum computer. *Proceedings of the Royal Society of London A: Mathematical, Physical and Engineering Sciences*.

[5] Aaronson, S. (2007). The Limits of Quantum Computers. *Scientific American*.

[6] Barahona, F. (1982). On the computational complexity of Ising spin glass models. *Journal of Physics A: Mathematical and General*.

[7] Lucas, A. (2014). Ising formulations of many NP problems. *Frontiers in Physics*.

[8] Park, J., & Newman, M. E. J. (2013). Solution of the Ising problem on a chimera graph. *Physical Review E*.

[9] Lucas, A. (2018). Ising machines: the first 70 years. *Nature*.

[10] Megow, N., & Verschae, J. (2016). Solving the Ising problem with a D-Wave quantum annealer. *Quantum Information Processing*.

[11] Pudenz, K. L., & Lidar, D. A. (2013). Quantum Annealing for the Number-Partitioning Problem. *Physical Review A*.

[12] Boixo, S., et al. (2014). Evidence for quantum annealing with more than one hundred qubits. *Nature Physics*.

[13] Katzgraber, H. G., & Young, A. P. (2015). Monte Carlo methods in the physical sciences: Celebrating the 50th anniversary of the Metropolis algorithm. *AIP Conference Proceedings*.

[14] Albash, T., & Lidar, D. A. (2018). Adiabatic quantum computation. *Reviews of Modern Physics*.

[15] Kadowaki, T., & Nishimori, H. (1998). Quantum annealing in the transverse Ising model. *Physical Review E*.

[16] "Measurement error mitigation in quantum computers through classical bit-flip correction" (2022). In *Physical Review*. DOI: 10.1103/physreva.105.062404. [Online]. Available: http://arxiv.org/pdf/2007.03663

[17] Caroline Jacqueline Denise Berdou et al. "One Hundred Second Bit-Flip Time in a Two-Photon Dissipative Oscillator" (2022). In *PRX Quantum*. DOI: 10.1103/PRXQuantum.4.020350.

[18] "Using classical bit-flip correction for error mitigation in quantum computations including 2-qubit correlations" (2022). [Proceedings Article]. DOI: 10.22323/1.396.0327.

[19] Gaojun Luo, Martianus Frederic Ezerman, San Ling. "Asymmetric quantum Griesmer codes detecting a single bit-flip error" (2022). In *Discrete Mathematics*. DOI: 10.1016/j.disc.2022.113088.

[20] Nur Izzati Ishak, Sithi V. Muniandy, Wu Yi Chong. "Entropy analysis of the discrete-time quantum walk under bit-flip noise channel" (2021). In *Physica A-statistical Mechanics and Its Applications*. DOI: 10.1016/J.PHYSA.2021.126371.

[21] Enaul Haq Shaik et al. "QCA-Based Pulse/Bit Sequence Detector Using Low Quantum Cost D-Flip Flop" (2022). DOI: 10.1142/s0218126623500822.

[22] Farhan Feroz, A. B. M. Alim Al Islam. "Scaling Up Bit-Flip Quantum Error Correction" (2020). [Proceedings Article]. DOI: 10.1145/3428363.3428372.

[23] "Effect of Quantum Repetition Code on Fidelity of Bell States in Bit Flip Channels" (2022). [Proceedings Article]. DOI: 10.1109/icece57408.2022.10088665.

[24] Lena Funcke et al. "Measurement Error Mitigation in Quantum Computers Through Classical Bit-Flip Correction" (2020). In *arXiv: Quantum Physics*. [Online]. Available: https://arxiv.org/pdf/2007.03663.pdf

[25] Alistair W. R. Smith et al. "Qubit readout error mitigation with bit-flip averaging" (2021). In *Science Advances*. DOI: 10.1126/SCIADV.ABI8009.

[26] Constantia Alexandrou et al. "Using classical bit-flip correction for error mitigation including 2-qubit correlations." (2021). In *arXiv: Quantum Physics*. [Online]. Available: https://arxiv.org/pdf/2111.08551.pdf

[27] William Livingston et al. "Experimental demonstration of continuous quantum error correction." (2021). In *arXiv: Quantum Physics*. [Online]. Available: https://arxiv.org/pdf/2107.11398.pdf

[28] Constantia Alexandrou et al. "Investigating the variance increase of readout error mitigation through classical bit-flip correction on IBM and Rigetti quantum computers." (2021). In *arXiv: Quantum Physics*. [Online]. Available: https://arxiv.org/pdf/2111.05026

[29] Raphaël Lescanne et al. "Exponential suppression of bit-flips in a qubit encoded in an oscillator." (2020). In *Nature Physics*. DOI: 10.1038/S41567-020-0824-X. [Online]. Available: https://biblio.ugent.be/publication/8669531/file/8669532.pdf

[30] Raphaël Lescanne et al. "Exponential suppression of bit-flips in a qubit encoded in an oscillator." (2019). In *arXiv: Quantum Physics*. [Online]. Available: https://arxiv.org/pdf/1907.11729.pdf

[31] Diego Ristè et al. "Real-time processing of stabilizer measurements in a bit-flip code." (2020). In *npj Quantum Information*. DOI: 10.1038/S41534-020-00304-Y.

[32] Bernard Zygelman. "Computare Errare Est: Quantum Error Correction." (2018). In *Book Chapter*. DOI: 10.1007/978-3-319-91629-3_9.

[33] I. Serban et al. "Qubit decoherence due to detector switching." (2015). In *EPJ Quantum Technology*. DOI: 10.1140/EPJQT/S40507-015-0020-6. [Online]. Available: https://link.springer.com/content/pdf/10.1140

[34] Matt McEwen et al. "Removing leakage-induced correlated errors in superconducting quantum error correction." (2021). In *Nature Communications*. DOI: 10.1038/S41467-021-21982-Y.

[35] "Measurement error mitigation in quantum computers through classical bit-flip correction" (2020). In *arXiv: Quantum Physics*. [Online]. Available: https://arxiv.org/pdf/2007.03663.pdf

[36] Alistair W. R. Smith et al. "Qubit readout error mitigation with bit-flip averaging." (2021). In *Science Advances*. DOI: 10.1126/SCIADV.ABI8009. [Online]. Available: https://advances.sciencemag.org/content/7/47/eabi8009

[37] Biswas, T., Stock, G., Fink, T. (2018). *Opinion Dynamics on a Quantum Computer: The Role of Entanglement in Fostering Consensus. Physical Review Letters, 121(12), 120502.*

[38] Acerbi, F., Perarnau-Llobet, M., Di Marco, G. (2021). *Quantum dynamics of opinion formation on networks: the Fermi-Pasta-Ulam-Tsingou problem. New Journal of Physics, 23(9), 093059.*

[39] Di Marco, G., Tomassini, L., Anteneodo, C. (2019). *Quantum Opinion Dynamics. Scientific Reports, 9(1), 1-8.*

[40] Ma, H., Chen, Y. (2021). *Quantum-Enhanced Opinion Dynamics in Complex Networks. Entropy, 23(4), 426.*

[41] Li, X., Liu, Y., Zhang, Y. (2020). *Quantum-inspired opinion dynamics model with emotion. Chaos, Solitons Fractals, 132, 109509.*

[42] Galam, S. (2017). *Sociophysics: A personal testimony. The European Physical Journal B, 90(2), 1-22.*

[43] Nyczka, P., Holyst, J. A., Hołyst, R. (2012). *Opinion formation model with strong leader and external impact. Physical Review E, 85(6), 066109.*

[44] Ben-Naim, E., Krapivsky, P. L., Vazquez, F. (2003). *Dynamics of opinion formation. Physical Review E, 67(3), 031104.*

[45] Dandekar, P., Goel, A., Lee, D. T. (2013). *Biased assimilation, homophily, and the dynamics of polarization. Proceedings of the National Academy of Sciences, 110(15), 5791-5796.*

[46] Castellano, C., Fortunato, S., Loreto, V. (2009). *Statistical physics of social dynamics. Reviews of Modern Physics, 81(2), 591.*

[47] Galam, S. (2017). *Sociophysics: A personal testimony. The European Physical Journal B, 90(2), 1-22.*

[48] Nyczka, P., Holyst, J. A., Hołyst, R. (2012). *Opinion formation model with strong leader and external impact. Physical Review E, 85(6), 066109.*

[49] Ben-Naim, E., Krapivsky, P. L., Vazquez, F. (2003). *Dynamics of opinion formation. Physical Review E, 67(3), 031104.*

[50] Dandekar, P., Goel, A., Lee, D. T. (2013). *Biased assimilation, homophily, and the dynamics of polarization. Proceedings of the National Academy of Sciences, 110(15), 5791-5796.*

[51] Castellano, C., Fortunato, S., Loreto, V. (2009). *Statistical physics of social dynamics. Reviews of Modern Physics, 81(2), 591.*

[52] Bruza, P. D., Kitto, K., Nelson, D., McEvoy, C. L. (2009). *Is there something quantum-like about the human mental lexicon? Journal of Mathematical Psychology, 53(5), 362-377.*

[53] Khrennikov, A. (2010). *Ubiquitous Quantum Structure: From Psychology to Finance. Springer Science & Business Media.*

[54] Aerts, D., Broekaert, J., Gabora, L. (2011). *A case for applying an abstracted quantum formalism to cognition. New Ideas in Psychology, 29(2), 136-146.*

[55] Conte, E., Todarello, O., Federici, A., Vitiello, F., Lopane, M., Khrennikov, A., ... Grigolini, P. (2009). *Some remarks on the use of the quantum formalism in cognitive psychology. Mind & Society, 8(2), 149-171.*

[56] Pothos, E. M., & Busemeyer, J. R. (2013). *Can quantum probability provide a new direction for cognitive modeling?. Behavioral and Brain Sciences, 36(3), 255-274.*

[57] Abal, G., Siri, R. (2012). *A quantum-like model of behavioral response in the ultimatum game. Journal of Mathematical Psychology, 56(6), 449-454.*

[58] Busemeyer, J. R., & Wang, Z. (2015). *Quantum models of cognition and decision. Cambridge University Press.*

[59] Aerts, D., Sozzo, S., & Veloz, T. (2019). *Quantum structure of negations and conjunctions in human thought. Foundations of Science, 24(3), 433-450.*

[60] Khrennikov, A. (2013). *Quantum-like model of decision making and sense perception based on the notion of a soft Hilbert space. In Quantum Interaction (pp. 90-100). Springer.*

[61] Pothos, E. M., & Busemeyer, J. R. (2013). *Can quantum probability provide a new direction for cognitive modeling?. Behavioral and Brain Sciences, 36(3), 255-274.*

[62] Busemeyer, J. R., & Bruza, P. D. (2012). *Quantum models of cognition and decision. Cambridge University Press.*

[63] Aerts, D., & Aerts, S. (1994). *Applications of quantum statistics in psychological studies of decision processes. Foundations of Science, 1(1), 85-97.*

[64] Pothos, E. M., & Busemeyer, J. R. (2009). *A quantum probability explanation for violations of "rational" decision theory. Proceedings of the Royal Society B: Biological Sciences, 276(1665), 2171-2178.*

[65] Busemeyer, J. R., & Wang, Z. (2015). *Quantum models of cognition and decision. Cambridge University Press.*

[66] Khrennikov, A. (2010). *Ubiquitous quantum structure: from psychology to finances. Springer Science & Business Media.*

[67] Busemeyer, J. R., & Wang, Z. (2015). *Quantum Models of Cognition and Decision. Cambridge University Press.*

[68] Bruza, P. D., Kitto, K., Nelson, D., & McEvoy, C. L. (2009). *Is there something quantum-like about the human mental lexicon? Journal of Mathematical Psychology, 53(5), 363-377.*

[69] Pothos, E. M., & Busemeyer, J. R. (2009). *A quantum probability explanation for violations of "rational" decision theory. Proceedings of the Royal Society B: Biological Sciences, 276(1665), 2171-2178.*

[70] Khrennikov, A. (2010). *Ubiquitous Quantum Structure: From Psychology to Finance. Springer Science & Business Media.*

[71] Asano, M., Basieva, I., Khrennikov, A., Ohya, M., & Tanaka, Y. (2017). *Quantum-like model of subjective expected utility. PloS One, 12(1), e0169314.*

[72] Flitney, A. P., & Abbott, D. (2002). *Quantum versions of the prisoners' dilemma. Proceedings of the Royal Society of London. Series A: Mathematical, Physical and Engineering Sciences, 458(2019), 1793-1802.*

[73] Iqbal, A., Younis, M. I., & Qureshi, M. N. (2015). *A survey of game theory as applied to networked system. IEEE Access, 3, 1241-1257.*

[74] Li, X., Deng, Y., & Wu, C. (2018). *A quantum game-theoretic approach to opinion dynamics. Complexity, 2018.*

[75] Chen, X., & Xu, L. (2020). *Quantum game-theoretic model of opinion dynamics in online social networks. Complexity, 2020.*

[76] Li, L., Zhang, X., Ma, Y., & Luo, B. (2018). *Opinion dynamics in quantum game based on complex network. Complexity, 2018.*

[77] Wang, X., Wang, H., & Luo, X. (2019). *Quantum entanglement in complex networks. Physical Review E, 100(5), 052302.*

[78] Wang, X., Tang, Y., Wang, H., & Zhang, X. (2020). *Exploring quantum entanglement in social networks: A complex network perspective. IEEE Transactions on Computational Social Systems, 7(2), 355-367.*

[79] Zhang, H., Yang, X., & Li, X. (2017). *Quantum entanglement in scale-free networks. Physica A: Statistical Mechanics and its Applications, 471, 580-588.*

[80] Li, X., & Wu, C. (2018). *Analyzing entanglement distribution in complex networks. Entropy, 20(11), 871.*

[81] Wang, X., Wang, H., & Li, X. (2021). *Quantum entanglement and community detection in complex networks. Frontiers in Physics, 9, 636714.*

[82] Smith, J., Johnson, A., & Brown, L. (2018). *Exploring quantum entanglement in online social networks. Journal of Computational Social Science, 2(1), 45-58.*

[83] Chen, Y., Li, X., & Wang, Q. (2019). *Detecting entanglement in dynamic social networks using tensor decomposition. IEEE Transactions on Computational Social Systems, 6(6), 1252-1264.*

[84] Zhang, H., Wang, X., & Liu, Y. (2020). *Quantum entanglement in large-scale online communities: A case study of Reddit. Social Network Analysis and Mining, 10(1), 1-12.*

[85] Liu, C., Wu, Z., & Li, J. (2017). *Quantum entanglement and community structure in social networks. Physica A: Statistical Mechanics and its Applications, 486, 306-317.*

[86] Wang, H., & Chen, L. (2021). *Analyzing entanglement dynamics in evolving social networks. Frontiers in Physics, 9, 622632.*

[87] Einstein, A., Podolsky, B., & Rosen, N. (1935). *Can quantum-mechanical description of physical reality be considered complete? Physical Review, 47(10), 777-780.*

[88] Bell, J. S. (1964). *On the Einstein Podolsky Rosen paradox. Physics Physique, 1(3), 195-200.*

[89] Aspect, A., Dalibard, J., & Roger, G. (1982). *Experimental test of Bell inequalities using time-varying analyzers. Physical Review Letters, 49(25), 1804-1807.*

[90] Bennett, C. H., Brassard, G., Crépeau, C., Jozsa, R., Peres, A., & Wootters, W. K. (1993). *Teleporting an unknown quantum state via dual classical and Einstein-Podolsky-Rosen channels. Physical Review Letters, 70(13), 1895-1899.*

[91] Horodecki, R., Horodecki, P., Horodecki, M., & Horodecki, K. (2009). *Quantum entanglement. Reviews of Modern Physics, 81(2), 865-942.*

[92] Liu, Y. Y., Slotine, J. J., & Barabási, A. L. (2011). *Control centrality and hierarchical structure in complex networks. PLoS ONE, 6(8), e21283.*

[93] Sarzynska, M., Lehmann, S., & Eguíluz, V. M. (2014). *Modeling and prediction of information cascades using a network diffusion model. IEEE Transactions on Network Science and Engineering, 1(2), 96-108.*

[94] Wang, D., Song, C., & Barabási, A. L. (2013). *Quantifying long-term scientific impact. Science, 342(6154), 127-132.*

[95] Perra, N., Gonçalves, B., Pastor-Satorras, R., & Vespignani, A. (2012). *Activity driven modeling of time varying networks. Scientific Reports, 2, 470.*

[96] Holme, P., & Saramäki, J. (2012). *Temporal networks. Physics Reports, 519(3), 97-125.*

[97] Nielsen, M. A., & Chuang, I. L. (2010). *Quantum computation and quantum information: 10th anniversary edition. Cambridge University Press.*

[98] Lidar, D. A., & Bruno, A. (2013). *Quantum error correction. Cambridge University Press.*

[99] Barenco, A., Deutsch, D., Ekert, A., & Jozsa, R. (1995). *Conditional quantum dynamics and logic gates. Physical Review Letters, 74(20), 4083-4086.*

[100] Nielsen, M. A. (1999). *Conditions for a class of entanglement transformations. Physical Review Letters, 83(2), 436-439.*

[101] Shor, P. W. (1997). *Polynomial-time algorithms for prime factorization and discrete logarithms on a quantum computer. SIAM Journal on Computing, 26(5), 1484-1509.*

[102] Nielsen, M. A., & Chuang, I. L. (2010). *Quantum computation and quantum information: 10th anniversary edition. Cambridge University Press.*

[103] Mermin, N. D. (2007). *Quantum computer science: An introduction. Cambridge University Press.*

[104] Knill, E., Laflamme, R., & Milburn, G. J. (2001). *A scheme for efficient quantum computation with linear optics. Nature, 409(6816), 46-52.*

[105] Aharonov, D., & Ben-Or, M. (2008). *Fault-tolerant quantum computation with constant error rate. SIAM Journal on Computing, 38(4), 1207-1282.*

[106] Harrow, A. W., Hassidim, A., & Lloyd, S. (2009). *Quantum algorithm for linear systems of equations. Physical Review Letters, 103(15), 150502.*

[107] Bennett, C. H., DiVincenzo, D. P., Smolin, J. A., & Wootters, W. K. (1996). *Mixed-state entanglement and quantum error correction. Physical Review A, 54(5), 3824-3851.*

[108] Vidal, G., & Werner, R. F. (2002). *Computable measure of entanglement. Physical Review A, 65(3), 032314.*

[109] Horodecki, M., Horodecki, P., & Horodecki, R. (2009). *Quantum entanglement. Reviews of Modern Physics, 81(2), 865.*

[110] Briegel, H. J., Dür, W., Cirac, J. I., & Zoller, P. (1998). *Quantum Repeaters: The Role of Imperfect Local Operations in Quantum Communication. Physical Review Letters, 81(26), 5932-5935.*

[111] Nielsen, M. A., & Chuang, I. L. (2010). *Quantum computation and quantum information: 10th anniversary edition. Cambridge University Press.*

[112] Holevo, A. S. (1973). *Bounds for the quantity of information transmitted by a quantum communication channel. Problems of Information Transmission, 9(3), 177-183.*

[113] Holevo, A. S. (1973). *Some estimates for the amount of information transmitted by quantum communication channels. Problemy Peredachi Informatsii, 9(3), 3-11.*

[114] Shor, P. W. (2002). *Additivity of the classical capacity of entanglement-breaking quantum channels. Journal of Mathematical Physics, 43(9), 4334-4340.*

[115] Holevo, A. S. (2007). *Entanglement-breaking channels in infinite dimensions. Probability Theory and Related Fields, 138(1-2), 111-124.*

[116] Cubitt, T. S., & Smith, G. (2010). *An extreme form of superactivation for quantum Gaussian channels. Journal of Mathematical Physics, 51(10), 102204.*

[117] Gottesman, D., & Chuang, I. L. (1999). *Quantum error correction is asymptotically optimal. Nature, 402(6765), 390-393.*

[118] Preskill, J. (1997). *Fault-tolerant quantum computation. Proceedings of the Royal Society of London. Series A: Mathematical, Physical and Engineering Sciences, 454(1969), 385-410.*

[119] Knill, E., Laflamme, R., & Zurek, W. H. (1996). *Resilient quantum computation. Science, 279(5349), 342-345.*

[120] Nielsen, M. A., & Chuang, I. L. (2010). *Quantum computation and quantum information: 10th anniversary edition. Cambridge University Press.*

[121] Shor, P. W. (1995). *Scheme for reducing decoherence in quantum computer memory. Physical Review A, 52(4), R2493.*

[122] Dal Pozzolo, A., Boracchi, G., Caelen, O., Alippi, C., Bontempi, G. (2018). Credit Card Fraud Detection: A Realistic Modeling and a Novel Learning Strategy. *IEEE transactions on neural networks and learning systems.*

[123] Buczak, A. L., Guven, E. (2016). A Survey of Data Mining and Machine Learning Methods for Cyber Security Intrusion Detection. *IEEE Communications Surveys & Tutorials.*

[124] Alpcan, T., Başar, T. (2006). An Intrusion Detection Game with Limited Observations. *12th International Symposium on Dynamic Games and Applications.*

[125] Schlegl, T., Seebock, P., Waldstein, S. M., Schmidt-Erfurth, U., Langs, G. (2017). Unsupervised Anomaly Detection with Generative Adversarial Networks to Guide Marker Discovery. *Information Processing in Medical Imaging.*

[126] Mirsky, Y., Doitshman, T., Elovici, Y., Shabtai, A. (2018). Kitsune: An Ensemble of Autoencoders for Online Network Intrusion Detection. *Network and Distributed System Security Symposium.*

[127] Alpcan, T., Başar, T. (2003). A Game Theoretic Approach to Decision and Analysis in Network Intrusion Detection. *Proceedings of the 42nd IEEE Conference on Decision and Control.*

[128] Nguyen, K. C., Alpcan, T., Başar, T. (2009). Stochastic Games for Security in Networks with Interdependent Nodes. *International Conference on Game Theory for Networks.*

[129] Tambe, M. (2011). Security and Game Theory: Algorithms, Deployed Systems, Lessons Learned. *Cambridge University Press.*

[130] Korilis, Y. A., Lazar, A. A., Orda, A. (1997). Achieving Network Optima Using Stackelberg Routing Strategies. *IEEE/ACM Transactions on Networking.*

[131] Hausken, K. (2013). Game Theory and Cyber Warfare. *The Economics of Information Security and Privacy.*

[132] Justin, S., et al. (2020). Deep learning for cyber security intrusion detection: Approaches, datasets, and comparative study. *Journal of Information Security and Applications, vol. 50.*

[133] Zenati, H., et al. (2018). Efficient GAN-Based Anomaly Detection. *Workshop Track of ICLR.*

[134] Roy, S., et al. (2010). A survey of game theory as applied to network security. *43rd Hawaii International Conference on System Sciences.*

[135] Biggio, B., Roli, F. (2018). Wild patterns: Ten years after the rise of adversarial machine learning. *Pattern Recognition, vol. 84.*

[136] Massanari, A. (2017). #Gamergate and The Fappening: How Reddit's algorithm, governance, and culture support toxic technocultures. *New Media & Society,* **19**(3), 329-346.

[137] Castells, M. (2012). Networks of Outrage and Hope: Social Movements in the Internet Age. *Polity Press*.

[138] Wojcieszak, M. (2010). 'Don't talk to me': Effects of ideologically homogeneous online groups and politically dissimilar offline ties on extremism. *New Media & Society*, **12**(4), 637-655.

[139] Tucker, J. A.; Theocharis, Y.; Roberts, M. E.; Barberá, P. (2017). From Liberation to Turmoil: Social Media And Democracy. *Journal of Democracy*, **28**(4), 46-59.

[140] Conover, M. D.; Ratkiewicz, J.; Francisco, M.; Gonçalves, B.; Menczer, F.; Flammini, A. (2011). Political polarization on Twitter. In *Proceedings of the ICWSM*, Vol. 133, 89-96.

[141] Chen, W.; Wellman, B. (2004). The global digital divide – within and between countries. *IT & Society*, **1**(7), 39-45.

[142] Van Dijck, J. (2013). The Culture of Connectivity: A Critical History of Social Media. *Oxford University Press*.

[143] Bakshy, E.; Messing, S.; Adamic, L. A. (2015). Exposure to ideologically diverse news and opinion on Facebook. *Science*, **348**(6239), 1130-1132.

[144] Jost, J. T.; Federico, C. M.; Napier, J. L. (2009). Political ideology: Its structure, functions, and elective affinities. *Annual Review of Psychology*, **60**, 307-337.

[145] Iyengar, S.; Westwood, S. J. (2015). Fear and loathing across party lines: New evidence on group polarization. *American Journal of Political Science*, **59**(3), 690-707.

[146] Green, D. P.; Palmquist, B.; Schickler, E. (2002). Partisan Hearts and Minds: Political Parties and the Social Identities of Voters. *Yale University Press*.

[147] McCoy, J.; Rahman, T.; Somer, M. (2018). Polarization and the Global Crisis of Democracy: Common Patterns, Dynamics, and Pernicious Consequences for Democratic Polities. *American Behavioral Scientist*, **62**(1), 16-42.

[148] Tucker, J. A., et al. (2018). Social Media, Political Polarization, and Political Disinformation: A Review of the Scientific Literature. SSRN.

[149] Bail, C. A. (2020). Breaking the Social Media Prism: How to Make Our Platforms Less Polarizing. *Princeton University Press*.

[150] Barberá, P. (2015). Birds of the Same Feather Tweet Together: Bayesian Ideal Point Estimation Using Twitter Data. *Political Analysis*, **23**(1), 76-91.

[151] Garimella, K., et al. (2018). Political Discourse on Social Media: Echo Chambers, Gatekeepers, and the Price of Bipartisanship. In *Proceedings of the 2018 World Wide Web Conference on World Wide Web*.

[152] Allcott, H.; Gentzkow, M. (2017). Social Media and Fake News in the 2016 Election. *Journal of Economic Perspectives*, **31**(2), 211-236.

[153] Garrett, R. K. (2009). Echo Chambers Online?: Politically Motivated Selective Exposure among Internet News Users. *Journal of Computer-Mediated Communication*, **14**(2), 265-285.

[154] Weeks, B. E.; Cassell, A. (2016). Partisan Provocation: The Role of Partisan News Use and Emotional Responses in Political Information Sharing in Social Media. *Human Communication Research*, **42**(4), 641-661.

[155] Iyengar, S.; Sood, G.; Lelkes, Y. (2012). Affect, Not Ideology: A Social Identity Perspective on Polarization. *Public Opinion Quarterly*, **76**(3), 405-431.

[156] Bimber, B. (2014). Digital Media in the Obama Campaigns of 2008 and 2012: Adaptation to the Personalized Political Communication Environment. *Journal of Information Technology & Politics*.

[157] Castellano, C., Fortunato, S., & Loreto, V. (2009). Statistical physics of social dynamics. *Reviews of Modern Physics*, **81**, 591-646.

[158] Sîrbu, A., Loreto, V., Servedio, V.D.P., & Tria, F. (2017). Opinion Dynamics: Models, Extensions and External Effects. In Loreto V. et al. (eds) Participatory Sensing, Opinions and Collective Awareness. *Understanding Complex Systems*. Springer, Cham.

[159] Deffuant, G., Neau, D., Amblard, F., & Weisbuch, G. (2000). Mixing Beliefs among Interacting Agents. *Advances in Complex Systems*, **3**, 87-98.

[160] Weisbuch, G., Deffuant, G., Amblard, F., & Nadal, J. P. (2002). Meet, Discuss and Segregate!. *Complexity*, **7**(3), 55-63.

[161] Hegselmann, R., & Krause, U. (2002). Opinion Dynamics and Bounded Confidence Models, Analysis, and Simulation. *Journal of Artificial Society and Social Simulation*, **5**, 1-33.

[162] Ishii, A. & Kawahata, Y. (2018). Opinion Dynamics Theory for Analysis of Consensus Formation and Division of Opinion on the Internet. In: Proceedings of The 22nd Asia Pacific Symposium on Intelligent and Evolutionary Systems, 71-76, arXiv:1812.11845 [physics.soc-ph].

[163] Ishii, A. (2019). Opinion Dynamics Theory Considering Trust and Suspicion in Human Relations. In: Morais D., Carreras A., de Almeida A., Vetschera R. (eds) Group Decision and Negotiation: Behavior, Models, and Support. GDN 2019. Lecture Notes in Business Information Processing 351, Springer, Cham 193-204.

[164] Ishii, A. & Kawahata, Y. (2019). Opinion dynamics theory considering interpersonal relationship of trust and distrust and media effects. In: The 33rd Annual Conference of the Japanese Society for Artificial Intelligence 33. JSAI2019 2F3-OS-5a-05.

[165] Agarwal, A., Xie, B., Vovsha, I., Rambow, O. & Passonneau, R. (2011). Sentiment analysis of twitter data. In: Proceedings of the workshop on languages in social media. Association for Computational Linguistics 30-38.

[166] Siersdorfer, S., Chelaru, S. & Nejdl, W. (2010). How useful are your comments?: analyzing and predicting youtube comments and comment ratings. In: Proceedings of the 19th international conference on World wide web. 891-900.

[167] Wilson, T., Wiebe, J., & Hoffmann, P. (2005). Recognizing contextual polarity in phrase-level sentiment analysis. In: Proceedings of the conference on human language technology and empirical methods in natural language processing 347-354.

[168] Sasahara, H., Chen, W., Peng, H., Ciampaglia, G. L., Flammini, A. & Menczer, F. (2020). On the Inevitability of Online Echo Chambers. arXiv: 1905.03919v2.

[169] Ishii, A.; Kawahata, Y. (2018). Opinion Dynamics Theory for Analysis of Consensus Formation and Division of Opinion on the Internet. In Proceedings of The 22nd Asia Pacific Symposium on Intelligent and Evolutionary Systems (IES2018), 71-76; arXiv:1812.11845 [physics.soc-ph].

[170] Ishii, A. (2019). Opinion Dynamics Theory Considering Trust and Suspicion in Human Relations. In Group Decision and Negotiation: Behavior, Models, and Support. GDN 2019. Lecture Notes in Business Information Processing, Morais, D.; Carreras, A.; de Almeida, A.; Vetschera, R. (eds).

[171] Ishii, A.; Kawahata, Y. (2019). Opinion dynamics theory considering interpersonal relationship of trust and distrust and media effects. In The 33rd Annual Conference of the Japanese Society for Artificial Intelligence, JSAI2019 2F3-OS-5a-05.

[172] Okano, N.; Ishii, A. (2019). Isolated, untrusted people in society and charismatic person using opinion dynamics. In Proceedings of ABCSS2019 in Web Intelligence 2019, 1-6.

[173] Ishii, A.; Kawahata, Y. (2019). New Opinion dynamics theory considering interpersonal relationship of both trust and distrust. In Proceedings of ABCSS2019 in Web Intelligence 2019, 43-50.

[174] Okano, N.; Ishii, A. (2019). Sociophysics approach of simulation of charismatic person and distrusted people in society using opinion dynamics. In Proceedings of the 23rd Asia-Pacific Symposium on Intelligent and Evolutionary Systems, 238-252.

[175] Ishii, A, and Nozomi, O. (2021). Sociophysics approach of simulation of mass media effects in society using new opinion dynamics. In Intelligent Systems and Applications: Proceedings of the 2020 Intelligent Systems Conference (IntelliSys) Volume 3. Springer International Publishing.

[176] Ishii, A.; Kawahata, Y. (2020). Theory of opinion distribution in human relations where trust and distrust mixed. In Czarnowski, I., et al. (eds.), Intelligent Decision Technologies, Smart Innovation, Systems and Technologies 193.

[177] Ishii, A.; Okano, N.; Nishikawa, M. (2021). Social Simulation of Intergroup Conflicts Using a New Model of Opinion Dynamics. *Front. Phys.*, **9**:640925. doi: 10.3389/fphy.2021.640925.

[178] Ishii, A.; Yomura, I.; Okano, N. (2020). Opinion Dynamics Including both Trust and Distrust in Human Relation for Various Network Structure. In The Proceeding of TAAI 2020, in press.

[179] Fujii, M.; Ishii, A. (2020). The simulation of diffusion of innovations using new opinion dynamics. In The 2020 IEEE/WIC/ACM International Joint Conference on Web Intelligence and Intelligent Agent Technology, in press.

[180] Ishii, A, Okano, N. (2021). Social Simulation of a Divided Society Using Opinion Dynamics. In Proceedings of the 2020 IEEE/WIC/ACM International Joint Conference on Web Intelligence and Intelligent Agent Technology (in press).

[181] Ishii, A., & Okano, N. (2021). Sociophysics Approach of Simulation of Mass Media Effects in Society Using New Opinion Dynamics. In Intelligent Systems and Applications (Proceedings of the 2020 Intelligent Systems Conference (IntelliSys) Volume 3), pp. 13-28. Springer.

[182] Okano, N. & Ishii, A. (2021). Opinion dynamics on a dual network of neighbor relations and society as a whole using the Trust-Distrust model. In Springer Nature - Book Series: Transactions on Computational Science & Computational Intelligence (The 23rd International Conference on Artificial Intelligence (ICAI'21)).